\documentclass[12pt]{article}

\usepackage{fullpage}
\usepackage{natbib}

\usepackage{acro}
\usepackage{amsmath}
\usepackage{amssymb}
\usepackage{amsthm}
\usepackage{bm}
\usepackage{booktabs}
\usepackage{graphicx} 
\usepackage{hyperref}
\usepackage{bbm}
\usepackage{cleveref}
\usepackage[usenames,dvipsnames]{xcolor}
\usepackage{xspace}
\usepackage{subcaption}

\DeclareAcronym{PrO}{short=PrO, long=predictively oriented}
\DeclareAcronym{EKF}{short=EKF, long=extended Kalman filter}
\DeclareAcronym{UKF}{short=UKF, long=unscented Kalman filter}
\DeclareAcronym{CKF}{short=CKF, long=cubature Kalman filter}
\DeclareAcronym{UT}{short=UT, long=unscented transform}
\DeclareAcronym{KLD}{short=KLD, long=Kullback--Leibler divergence}
\DeclareAcronym{DoC}{short=DoC, long=difference-of-convex}
\DeclareAcronym{MSE}{short=MSE, long=mean square error}
\DeclareAcronym{NLL}{short=NLL, long=negative log-likelihood}

\definecolor{EKF}{RGB}{32, 178, 170}
\definecolor{IEKF}{RGB}{100, 200, 100}
\definecolor{EKF-B}{RGB}{255, 140, 0}
\definecolor{EKF-IW}{RGB}{220, 20, 60}
\definecolor{OGD}{RGB}{138, 43, 226}
\definecolor{WLF-IMQ}{RGB}{30, 144, 255}
\definecolor{WLF-MD}{RGB} {229, 193, 0}
\definecolor{Hub-EnKF}{RGB}{0, 116, 0}
\definecolor{WS-EnKF}{RGB}{198, 122, 55}
\definecolor{tabblue}{RGB}{31, 119, 180}
\definecolor{taborange}{RGB}{255, 127, 14}
\definecolor{mediumorchid}{HTML}{BA55D3}

\definecolor{CKF}{RGB}{123, 104, 238}
\definecolor{CKF-PrO}{RGB}{184, 134, 11}
\definecolor{CKF-PrO-Full}{RGB}{102, 74, 6}

\definecolor{UKF-plain}{RGB}{0, 128, 128}
\definecolor{UKF-PrO}{RGB}{255, 20, 147}
\definecolor{UKF-PrO-Full}{RGB}{139, 0, 84}

\newcommand{\mWlfMd}{{\color{WLF-MD} \texttt{\textbf{WoLF-TMD}}}\xspace} %
\newcommand{\mWlfImq}{{\color{WLF-IMQ} \texttt{\textbf{WoLF-IMQ}}}\xspace}

\newcommand{\mkf}{{\color{EKF} \texttt{\textbf{KF}}}\xspace}
\newcommand{\mkfExtended}{{\color{EKF} \texttt{\textbf{EKF}}}\xspace}

\newcommand{\mAgamenoniExtended}{{\color{EKF-IW} \texttt{\textbf{EKF-IW}}}\xspace}
\newcommand{\mWangExtended}{{\color{EKF-B} \texttt{\textbf{EKF-B}}}\xspace}
\newcommand{\mPrO}{{\color{mediumorchid}\texttt{\textbf{EKF-PrO}}}\xspace}
\newcommand{\mPrOKF}{{\color{mediumorchid}\texttt{\textbf{KF-PrO}}}\xspace}
\newcommand{\mUKFPrO}{{\color{UKF-PrO}\texttt{\textbf{UKF-PrO}}}\xspace}
\newcommand{\mUKF}{{\color{UKF-plain}\texttt{\textbf{UKF}}}\xspace}

\newcommand{\mCKF}{{\color{CKF}\texttt{\textbf{CKF}}}\xspace}
\newcommand{\mCKFPrO}{{\color{CKF-PrO}\texttt{\textbf{CKF-PrO}}}\xspace}

\newcommand{\mUKFPrOFull}{{\color{UKF-PrO-Full}\texttt{\textbf{UKF-PrO}${}^{+}$}}\xspace}
\newcommand{\mCKFPrOFull}{{\color{CKF-PrO-Full}\texttt{\textbf{CKF-PrO}${}^{+}$}}\xspace}

\DeclareMathOperator*{\argmin}{arg\,min}

\newcommand{\dimlatent}{d}

\newtheorem{remark}{Remark}

\newtheorem{definition}{Definition}

\usepackage{algorithm}
\usepackage{algpseudocode}

\algrenewcommand\algorithmiccomment[1]{\hfill$\triangleright$ #1}
\algrenewcommand\algorithmicrequire{\textbf{Input:}}
\algrenewcommand\algorithmicensure{\textbf{Output:}}

\usepackage{tikz}
\usetikzlibrary{arrows.meta}
\usetikzlibrary{positioning}

\newif\ifshowextra

\showextrafalse

\title{An Unscented PrO Kalman Filter}
\author{Dan Kean, Zheyang Shen, Chris. J. Oates \\
\small Newcastle University 
}

\begin{document}

\maketitle

\begin{abstract}
Predictively-oriented (PrO) filtering for state-space models was proposed in \citet{shen2026predictively} as a promising route to avoiding over-confident inferences in settings where either the dynamical model, the measurement model, or both, could be misspecified.
Our focus is on nonlinear state-space models, where the local linearisation used to construct the PrO filter in \citet{shen2026predictively} deteriorates as nonlinearities enter either the dynamical model or the measurement model.
Our proposed solution is based on \emph{unscented transforms}, an established tool from classical filtering which provides a more accurate mechanism to propagate Gaussian uncertainty through a nonlinear functional, and which includes the cubature rule as a special case.
We introduce two variants: replacing the linearisation in the predict step alone, which we call \mUKFPrO, has a computational cost comparable to the existing PrO filter; replacing it in the update step as well, \mUKFPrOFull, requires a substantially more expensive joint optimisation.
Performance is empirically assessed on the Lorenz96 model under conditions of increasing difficulty, in which the state-space model is systematically misspecified; we find that \mUKFPrO{} with the parameters of the cubature Kalman filter offers consistent improvements, which become substantial under strong nonlinearity, whereas the default unscented parameters are unreliable, and that the full update-step extension with the same parameters does not justify its additional cost.
\end{abstract}

\section{Introduction}
\label{sec: intro}

Our setting is a state-space model
\begin{align}
    x_k & = f(x_{k-1} , \epsilon_k) \label{eq: SS1} \\
    y_k & = h(x_k , \xi_k) \label{eq: SS2}
\end{align}
where the noise terms $(\epsilon_i)_{i \in \mathbb{N}}$ and $(\xi_i)_{i \in \mathbb{N}}$ are independent.
The functions $f$ and $h$ may be nonlinear and only the $y_k$ are observed.
The \emph{filtering} task refers to the inference of the state variable $x_k$ given the observations $y_{1:k}:=(y_i)_{i=1}^k$.
Note that this set-up also allows for $f$ and $h$ to depend on one or more unknown (static) parameters $\theta$, which can be cast as additional components of the state variable $x_k$ and held constant under the dynamic model \eqref{eq: SS1}.
State-space models of the form \eqref{eq: SS1} and \eqref{eq: SS2} arise in time series analysis \citep{durbin2012time}, tracking in computational vision \citep{chen2011kalman}, and sensor fusion in robotics \citep{thrun2002probabilistic}; in all of these settings, $f$ and/or $h$ are frequently \emph{nonlinear}.

Since the latent state $x_k$ may not be strongly constrained from the observations $y_{1:k}$, a Bayesian approach to quantify uncertainty in $x_k$ is natural in the filtering context.
For linear $f$, $h$, and Gaussian $\epsilon_k$, $\xi_k$ with known mean and covariance, the \emph{Kalman filter} (\mkf)
is an efficient online algorithm for exact calculation of the posterior (or \emph{filtering}) distributions $p(x_k| y_{1:k})$ \citep{kalman1960}.
For nonlinear $f$, $h$, algorithms such as the \emph{extended} Kalman filter (\mkfExtended) exploit local linearisation to approximate the filtering distribution while retaining the efficient online nature of the \mkf, while algorithms such as the \emph{unscented} Kalman filter (\mUKF; \citealp{julier1996general}) instead propagate a deterministic set of \emph{sigma points} through $f$ and $h$, avoiding an explicit linearisation step and typically providing a more accurate approximation when nonlinearities are pronounced \citep{sarkka2007unscented}.
A textbook treatment of algorithms for Bayesian inference in the state-space model in \eqref{eq: SS1} and \eqref{eq: SS2} can be found in \citet{Sarkka23}.

The focus of this paper is the  filtering problem in settings where
\emph{the true data-generating process is not an instance of the state-space model} in \eqref{eq: SS1} and \eqref{eq: SS2}.
In such settings Bayesian inference, and the various numerical methods that approximate Bayesian inference (e.g. \mkfExtended), can produce inferences and predictions that are \emph{over-confident} because they are not capable of recognising and adapting to model misspecification.
Misspecification of the measurement model, most commonly through outliers \citep{liu2020data}, has been addressed using hierarchical measurement models \citep{ting2007, piche2012, Agamennoni2012, nurminen2015, Huang2016, wang2018} and robust loss functions \citep{boncelet1983, karlgaard2015nonlinear, das2023robust}; the latter are \emph{generalised} Bayesian methods \citep{bissiri2016general,knoblauch2019generalized}, developed for filtering by \citet{boustati2020generalised,duran2024outlier,duran2025unifying}, which down-weight observations far into the tail of the predictive distribution.
These protect against occasional outliers but not against \emph{systematic} misspecification; if all data were deemed to be outliers, then all data would be ignored.
Misspecification of the dynamical model, for example through manoeuvres or unmodelled phase transitions, can instead cause \emph{divergence} \citep{schlee1967divergencekf,fitzgerald2003divergencekf}.
Remedies include heavy-tailed noise models \citep{roth2013studentfilter,huang2017ststfilter}, variational estimation of the dynamics \citep{roth2017studentfilter,huang2020slidekf,vilmarest2024viking} and covariance inflation \citep{mehra2003approaches,kelly2002ridgekf,chang2023low}, but these require \emph{a priori} knowledge of how the model is misspecified, or additional hyper-parameters, as does learning the dynamics \citep[e.g.][]{dahan2025bayesian}.
More fundamentally, (generalised) Bayesian posteriors do not provide calibrated uncertainty when the statistical model is misspecified \citep{shenprediction}; in the filtering context, this collapse of uncertainty causes over-confident inferences and predictions.

\subsection{Our Contributions}

To deal with generic model misspecification in \eqref{eq: SS1} and \eqref{eq: SS2}, we contend that an alternative to (generalised) Bayesian inference is required; rather than attempting to fit a more accurate model in a Bayesian framework in the hope that the resulting posterior predictions will be well-calibrated, we target instead the \emph{predictive} performance of the original state-space model.
This is the perspective taken by \citet{shenprediction,lai2024predictive,mclatchie2025predictively}, who introduced \emph{predictively oriented} (\acs{PrO}) posteriors as a post-Bayesian alternative that seeks a distribution over latent variables for which the implied data predictive distribution most closely matches the observed dataset, and by \citet{shen2026predictively}, which developed this idea into a filtering algorithm for state-space models.
\Cref{sec: background} recaps the relevant material from these earlier works: the \ac{PrO} filtering objective, its associated \ac{KLD}-minimisation interpretation \eqref{eq: min KL}, and the resulting \mPrO \citep{shen2026predictively}, a fast, hyper-parameter-free approximate update obtained by linearising $f$ and $h$ locally --- analogous to how the \mkfExtended approximates the \mkf for nonlinear models.

The present paper is concerned with a specific limitation of \mPrO: as with the \mkfExtended itself, the local-linearisation heuristic on which \mPrO relies deteriorates as nonlinearities in $f$ and/or $h$ become more pronounced, since a single Jacobian evaluation can be a poor description of $f$ or $h$ over the effective support of the current filtering or predictive distribution.
In this regime the classical \mkfExtended is well known to be outperformed by \emph{sigma-point} methods such as the \mUKF \citep{julier1996general,sarkka2007unscented}, which propagate uncertainty through nonlinear functionals more accurately than a single local linearisation, at a computational cost that remains comparable to the \mkfExtended. 
Our contribution is to investigate the analogous enhancement for \ac{PrO} filtering, and to show that the benefit depends on \emph{where} the sigma point methods are applied:
\begin{enumerate}
    \item In \Cref{subsec: PrO UKF prediction} we develop \mUKFPrO, which replaces the Jacobian-based linearisation used by \mPrO with an approximation based on the \ac{UT} in the \emph{prediction} step alone. This substitution requires no change to \mPrO's update-step machinery, and reduces exactly to \mPrO in the linear regime where linearisation is already exact.
    \item In \Cref{subsec: full ukf pro} we develop the analogous replacement in the \emph{update} step as well, \mUKFPrOFull, and show that -- unlike the predict-step substitution -- this cannot proceed by direct substitution: it requires abandoning \mPrO's closed-form profiled-mean update in favour of a joint optimisation over the filtering mean and covariance.
    \item Through detailed empirical assessment in \Cref{sec: experiments}, we demonstrate that \mUKFPrO{} with the parameters of the cubature Kalman filter, which we denote \mCKFPrO, matches or out-performs \mPrO in every condition considered, and substantially so once nonlinearity in the dynamical model becomes pronounced, whereas \mUKFPrO{} with the default \ac{UT} parameters is unreliable (\Cref{subsubsec: lorenz-oscillating}); we further show that the corresponding full update-step extension, \mCKFPrOFull, offers no consistent additional benefit over \mCKFPrO{} when $h$ is nonlinear, at approximately $7\times$ the computational cost, and report this as a negative finding of independent interest.
\end{enumerate}
A discussion of our findings, and a recommendation for which construction to adopt in practice, is contained in \Cref{sec: discuss}.

\section{Background: Predictively-Oriented Filtering}
\label{sec: background}

We recap the material from \citet{shen2026predictively} that we build on; full derivations and proofs are contained there.

\subsection{Filtering as an Optimisation Task}
\label{subsec: Op-cen filter}

The latent state $x_k$ is assumed to take values in $\mathbb{R}^d$ and we let $\mathcal{P}(\mathbb{R}^d)$ denote the set of probability distributions on $\mathbb{R}^d$.
The conditional distribution of $x_k$ (resp. $y_k$) given $x_{k-1}$ implied by \eqref{eq: SS1} and \eqref{eq: SS2} is denoted $p(x_k | x_{k-1})$ (resp. $p(y_k | x_k)$), and a prediction for $x_k$ given the observations $y_{1:l}$ is denoted $q_{k|l} \in \mathcal{P}(\mathbb{R}^d)$, where to improve presentation we interchange freely between densities and distributions, and use the standard filtering notation in \citet{Sarkka23}.

Existing Bayesian and generalised Bayesian filtering methods can be formulated as minimising objective functionals defined on $\mathcal{P}(\mathbb{R}^d)$.
Following standard filtering terminology, this is done in two steps: the \emph{prediction} step uses the filtering distribution from the previous step $q_{k-1|k-1}$ to calculate the latent predictive distribution $q_{k|k-1}$, then the \emph{update} step minimises an objective function in which $q_{k|k-1}$ appears as a regulariser via the \ac{KLD}:

\paragraph{Prediction Step}

Given the filtering distribution from the previous time step, $q_{k-1|k-1}(x_{k-1})$, the latent predictive distribution is obtained using the Chapman--Kolmogorov equation
\begin{align}
	q_{k|k-1}(x_k) := \int p(x_k | x_{k-1}) q_{k-1|k-1}(x_{k-1})\mathrm{d} x_{k-1}.  \label{eq: CK}
\end{align}

\paragraph{Update Step}

Given the latent predictive distribution in \eqref{eq: CK}, the filtering distribution is formulated as the solution of an infinite-dimensional optimisation problem on the set of probability measures
\begin{align}
	q_{k|k} := \argmin_{q \in \mathcal{P}(\mathbb{R}^d)} \quad \ell (y_k; q) + \mathrm{KL}(q \Vert q_{k|k-1}) \label{eq: prob filter}
\end{align}
where the loss function $\ell$ is to be specified.

\bigskip

\noindent
This formulation unifies Bayesian and generalised Bayesian filtering: Bayesian inference (and, for linear $f,h$ and Gaussian noise, the \mkf) is recovered with the loss
\begin{align}
\ell^{\mathrm{(Bayes)}}(y_k; q):=-\int \log (p(y_k | x_k)) q(x_k)\mathrm{d}x_k \label{eq: Bayes loss}
\end{align}
\citep{knoblauch2019generalized}, while generalised Bayesian methods such as WoLF \citep{duran2024outlier} re-weight it.
Given sufficient data, however, both collapse onto a single ``best'' parameter \citep{miller2021asymptotic}, giving arbitrarily confident predictions even when the state-space model is misspecified.

\subsection{The Predictively-Oriented Filtering Objective}
\label{subsec: Pro GF}

Taking inspiration from the earlier work of \citet{masegosa2020learning,jankowiak2020parametric,sheth2020pseudo,morningstar2022pacm,lai2024predictive,shenprediction,mclatchie2025predictively,liu2025detecting} in the static inference context, \citet{shen2026predictively} define the \ac{PrO} filtering objective as follows:

\begin{definition}[\ac{PrO} filtering objective; \citealp{shen2026predictively}]
    The \emph{\ac{PrO} filter} employs the loss
    \begin{align}
    \ell^{(\mathrm{PrO})}(y_k; q) :=-\log \left(\int p(y_k | x_k) q(x_k)\mathrm{d}x_k\right) . \label{eq: pro loss}
    \end{align}
\end{definition}

\noindent Compared to the Bayesian loss \eqref{eq: Bayes loss}, the order of the logarithm and the integral are interchanged; viewing $q$ (rather than $x_k$) as the `parameter' of the model performs a probabilistic \emph{lifting} of the original measurement model, $p_q(y_k) := \int p(y_k | x_k) q(x_k)\mathrm{d}x_k$, in a similar manner to nonparametric maximum likelihood \citep{laird1978nonparametric}, and does not introduce any hyper-parameter which needs to be specified.
This key difference causes the \ac{PrO} filter to behave in a more desirable way compared to (generalised) Bayesian methods when the state-space model is misspecified: considering $y_k$ to be a sample from the true data-generating distribution $p_{\text{true}}$,
\begin{align}
	\mathbb{E}_{y_k\sim p_{\text{true}}(\cdot | x_k)}[\ell^{(\text{PrO})}(y_k ; q)] & = \underbrace{\mathrm{KL}\left(p_{\text{true}}(\cdot | x_k) \left\Vert p_q \right.\right)}_{\text{predictive fit}} \quad + \; \mathrm{constant} , \label{eq: min KL}
\end{align}
which demonstrates that \ac{PrO} filtering explicitly minimises the \ac{KLD} between the predictive $p_q$ and the true data distribution $p_{\text{true}}(\cdot | x_k)$, automatically calibrating predictions even when the model is misspecified \citep{mclatchie2025predictively}.

Exact evaluation of the update \eqref{eq: prob filter} under \eqref{eq: pro loss} is intractable, and general-purpose approximations \citep{shenprediction,mclatchie2025predictively,chazal2025computable} are too costly for filtering; \citet{shen2026predictively} instead develop a fast update based on recursive Gaussian variational approximations~\citep{lambert2022recursive, jones2024bayesian}, which we recap next.

\subsection{The Linear-Gaussian Case: PrO Kalman Filtering}
\label{sec: kalman compare}

Consider first the case where $f$, $h$ are linear and $\epsilon_k$, $\xi_k$ are Gaussian, i.e. the setting of the \mkf:
\begin{align*}
    x_k | x_{k-1} & \sim \mathcal{N}(A_{k-1}x_{k-1}, \Lambda_{k-1}) \\
	y_k | x_k & \sim \mathcal{N}(H_k x_k, R_k)
\end{align*}
for positive definite $\Lambda_{k-1}$ and $R_k$.
The optimal filtering distribution $q$ minimising the \ac{PrO} objective~\eqref{eq: prob filter} is not necessarily Gaussian, even with a Gaussian prior $q_{k|k-1}$; \citet{shen2026predictively} therefore define the \mPrOKF as the optimal filtering distribution within the Gaussian variational family

\begin{equation}
	q_{k|k} := \argmin_{q \in \mathcal{Q}(\mathbb{R}^d)} \Big\{ -\log \Big(\int p(y_k | x_k)\, q(x_k)\,\mathrm{d}x_k\Big) + \mathrm{KL}(q \Vert q_{k|k-1}) \Big\}, \label{eq: pro opt task}
\end{equation}

where $\mathcal{Q}(\mathbb{R}^d)$ is the set of all Gaussian distributions on $\mathbb{R}^d$.

Given a filtering distribution $q_{k-1|k-1} = \mathcal{N}(m_{k-1}, P_{k-1})$ from the previous time step, the \textbf{prediction step} coincides with the standard \mkf, since \eqref{eq: CK} has the closed-form solution
\begin{align}
    q_{k|k-1} &= \mathcal{N}(\tilde{m}_k, \tilde{P}_k), \qquad
	\tilde{m}_k = A_{k-1} m_{k-1}, \nonumber\\
    \tilde{P}_k &= A_{k-1}P_{k-1}A_{k-1}^\top + \Lambda_{k-1}.  \label{eq: predictive step}
\end{align}
The \textbf{update step}, however, does not have a closed form in general.
\citet{shen2026predictively} instead work directly with the objective in \eqref{eq: pro opt task}, parametrised as $q = \mathcal{N}(m, P)$:

\begin{equation}
	\mathcal{J}_k(m, P) \stackrel{+C}{:=} \frac{1}{2}\Big[ \mathrm{tr}\big(\tilde{P}_k^{-1} P\big) + \big\Vert \tilde{P}_k^{-\frac{1}{2}} \left(m - \tilde{m}_k\right)\big\Vert^2 + \log \frac{|S_k(P)|}{|P|} + \big\Vert S_k^{-\frac{1}{2}}(P)\, v_k(m)\big\Vert^2\Big] \label{eq: pro_gaussian}
\end{equation}

where $S_k(P):= H_k PH_k^\top + R_k$ and $v_k(m):= y_k-H_k m$ are the predictive covariance and innovation implied by $q=\mathcal{N}(m,P)$.
\citet{shen2026predictively} minimise $\mathcal{J}_k$ over $m$ in closed form and the resulting profiled objective over $P$ numerically, giving $(m_k,P_k)$.
This defines $q_{k|k} = \mathcal{N}(m_k,P_k)$ and completes the iterative specification of the \mPrOKF.

\subsection{Limitations of Local Linearisation}

Preserving the tractability of the \mPrOKF for general nonlinear filtering tasks inspired the local-linearisation approach taken in \citet{shen2026predictively}; for completeness, full details are contained in \Cref{app: EKF PrO detail}.
However, for \emph{strongly nonlinear} systems we will see that their \mPrO can fail.
The issue, as with the ordinary \mkfExtended, is that a single Jacobian evaluation can be a poor local description of $f$ or $h$ once nonlinearities become pronounced, which is exactly the setting in which \mkfExtended is known to be outperformed by sigma-point methods such as the \mUKF \citep{julier1996general,morelande2013analysis}.

\section{A PrO Unscented Kalman Filter}
\label{sec: pro ukf}

We now develop \mUKFPrO{} (\Cref{subsec: PrO UKF prediction}) and \mUKFPrOFull{} (\Cref{subsec: full ukf pro}).

\subsection{The Unscented Transform}
\label{subsec: UT recap}

We first recall the \ac{UT} \citep{julier1996general,sarkka2007unscented}, which approximates the mean and covariance of $g(z)$ for $z \sim \mathcal{N}(\mu, \Sigma)$, $z \in \mathbb{R}^d$, and a (possibly nonlinear) map $g$, without evaluating any derivatives of $g$.
A set of $2d+1$ \emph{sigma points} $\{\mathcal{Z}^{(i)}\}_{i=0}^{2d}$ and weights $\{w_m^{(i)}, w_c^{(i)}\}_{i=0}^{2d}$ is constructed deterministically from $(\mu, \Sigma)$,

\begin{align*}
    \mathcal{Z}^{(0)} &= \mu, 
        & w_m^{(0)} &= \tfrac{\lambda}{d+\lambda}, \\
    \mathcal{Z}^{(i)} &= \mu + \big[\sqrt{(d+\lambda)\Sigma}\big]_i, \quad i=1,\dots,d,
        & w_c^{(0)} &= \tfrac{\lambda}{d+\lambda} + (1-\alpha^2+\beta), \\
    \mathcal{Z}^{(i+d)} &= \mu - \big[\sqrt{(d+\lambda)\Sigma}\big]_i, \quad i=1,\dots,d,
        & w_m^{(i)} &= w_c^{(i)} = \tfrac{1}{2(d+\lambda)}, \quad i=1,\dots,2d.
\end{align*}
where $\lambda = \alpha^2(d+\kappa) - d$, $[\sqrt{\Sigma}]_i$ denotes the $i$th column of a matrix square root of $\Sigma$, and $(\alpha, \beta, \kappa)$ are constants for which standard default choices exist (e.g.\ $\alpha=10^{-3}$, $\beta=2$, $\kappa=0$; \citealp{Sarkka23}). These constants are usually fixed rather than tuned, but, as we show in \Cref{sec: experiments}, the results can depend strongly on them.
A particular special case is worth highlighting: setting $\alpha = 1$, $\beta = 0$, $\kappa = 0$ recovers the \ac{CKF} \citep{arasaratnam2009}, with $2\dimlatent$ sigma points at $\mu \pm \sqrt{\dimlatent}\,[\sqrt{\Sigma}]_i$ and equal weights $1/(2\dimlatent)$ (the centre point receiving zero weight). The \ac{CKF} arises from a parameter-free construction (spherical-radial cubature integration), and avoids the large negative centre weight of the default setting ($w_m^{(0)} \approx -10^{6}$ for $\dimlatent=60$), a known source of numerical difficulty for the \ac{UKF} in higher dimensions \citep{arasaratnam2009}. We refer to \mUKFPrO{} and \mUKFPrOFull{} with these parameters as \mCKFPrO{} and \mCKFPrOFull{} respectively. Throughout, we fix $\alpha=10^{-3}$, $\beta=2$, $\kappa=0$ for \mUKFPrO/\mUKFPrOFull, and $\alpha=1,\beta=0,\kappa=0$ (exactly) for \mCKFPrO/\mCKFPrOFull, never tuned per experiment.
Propagating each sigma point through $g$, and letting $\mathcal{G}^{(i)} := g(\mathcal{Z}^{(i)})$, the \ac{UT} approximates
\begin{align}
    \mathbb{E}[g(z)] &\approx \bar{\mathcal{G}} := \textstyle\sum_{i=0}^{2d} w_m^{(i)} \mathcal{G}^{(i)}, \nonumber\\
    \mathrm{Cov}[g(z)] &\approx \textstyle\sum_{i=0}^{2d} w_c^{(i)} \big(\mathcal{G}^{(i)} - \bar{\mathcal{G}}\big)\big(\mathcal{G}^{(i)} - \bar{\mathcal{G}}\big)^\top . \label{eq: UT moments}
\end{align}
Unlike local linearisation, \eqref{eq: UT moments} 
captures curvature information to (at least) second order in $g$ around $\mu$ without requiring $g$ to be differentiable \citep{julier1996general}.

\subsection{Prediction Step via the Unscented Transform: \mUKFPrO}
\label{subsec: PrO UKF prediction}

Given $q_{k-1|k-1} = \mathcal{N}(m_{k-1}, P_{k-1})$, we introduce 
a prediction step inspired by the \ac{UKF} \citep[Algorithm~5.14]{Sarkka23}.
To this end, we form sigma points $\{\mathcal{X}^{(i)}\}_{i=0}^{2d}$ from $(m_{k-1}, P_{k-1})$ as in \Cref{subsec: UT recap}, propagate $\hat{\mathcal{X}}^{(i)} := f(\mathcal{X}^{(i)}, 0)$, and set
\begin{align}
    \tilde m_k &= \textstyle\sum_{i=0}^{2d} w_m^{(i)} \hat{\mathcal{X}}^{(i)}, \nonumber\\
    \tilde P_k &= \textstyle\sum_{i=0}^{2d} w_c^{(i)} \big(\hat{\mathcal{X}}^{(i)} - \tilde m_k\big)\big(\hat{\mathcal{X}}^{(i)} - \tilde m_k\big)^\top + \Lambda_{k-1} , \label{eq: ukf predictive step}
\end{align}
where we have assumed additive process noise, $f(x, \epsilon) = f(x, 0) + \epsilon$ with $\epsilon \sim \mathcal{N}(0, \Lambda_{k-1})$, as in all our experiments; non-additive noise can be handled by augmenting the state with $\epsilon$ in the usual way \citep{Sarkka23}.
The resulting prediction is $q_{k|k-1} = \mathcal{N}(\tilde m_k, \tilde P_k)$.
Since this has the same Gaussian form as \Cref{sec: kalman compare}, one can immediately employ the same update step as the \mPrOKF (or the \mPrO, if one is prepared to locally linearise the observational process $h$); we call this method \mUKFPrO, and refer to the special case using the cubature parameters of \Cref{subsec: UT recap} as \mCKFPrO.
However, in settings where \emph{both} the dynamical process $f$ and the observation process $h$ are strongly nonlinear, an alternative update might be needed; we turn to this next.

\subsection{Update Step via the Unscented Transform: \mUKFPrOFull}
\label{subsec: full ukf pro}

To proceed using sigma points also for the observation process an alternative update step is required; we will recompute the moment-matched approximation directly as a function of $(m,P)$, rather than linearising once around a fixed reference point.

The predictive density implied by a candidate $q=\mathcal{N}(m,P)$, $p_q(y_k) = \int p(y_k | x_k)\,\mathcal{N}(x_k; m, P)\,\mathrm{d}x_k$, is the distribution of $h(x_k,0)+\xi_k$ with $x_k\sim\mathcal{N}(m,P)$ and independent $\xi_k\sim\mathcal{N}(0,R_k)$, so that $\mathbb{E}[y_k]=\mathbb{E}[h(x_k,0)]$ and $\mathrm{Cov}[y_k]=\mathrm{Cov}[h(x_k,0)]+R_k$, neither of which admits a closed form for nonlinear $h$. Forming sigma points $\{\mathcal{X}^{(i)}(m,P)\}_{i=0}^{2d}$ from $(m,P)$ (\Cref{subsec: UT recap}) and $\mathcal{Y}^{(i)}(m,P) := h(\mathcal{X}^{(i)}(m,P),0)$, we approximate these moments using the same sigma-point rule as in the prediction step:
\begin{align}
    \mu(m,P) &:= \textstyle\sum_i w_m^{(i)} \mathcal{Y}^{(i)}, \label{eq: ukf mu} \\
    \Sigma(m,P) &:= \textstyle\sum_i w_c^{(i)} \big(\mathcal{Y}^{(i)} - \mu\big)\big(\mathcal{Y}^{(i)} - \mu\big)^\top + R_k, \label{eq: ukf Sigma}
\end{align}
where we suppress the dependence of $\mathcal{Y}^{(i)}$ and $\mu$ on $(m,P)$.
Approximating $p_q(y_k)$ by the Gaussian sharing these moments, 
the loss becomes
\begin{equation}
    \ell^{(\mathrm{PrO})}(y_k;q) \approx -\log N\big(y_k;\mu(m,P),\Sigma(m,P)\big), \label{eq: ukf loss}
\end{equation}
with the approximation being exact for linear $h$.
The loss \eqref{eq: ukf loss} is an explicit function of $(m,P)$, since the sigma points \eqref{eq: ukf mu}--\eqref{eq: ukf Sigma} are re-formed at every candidate $(m,P)$, rather than linearised once around a fixed reference point.

Plugging \eqref{eq: ukf loss} into \eqref{eq: pro_gaussian} in place of the corresponding term derived from \eqref{eq: pro linear1}--\eqref{eq: pro linear2}, together with the predict-step substitution of \Cref{subsec: PrO UKF prediction}, defines \mUKFPrOFull{} (and, with the cubature parameters of \Cref{subsec: UT recap}, \mCKFPrOFull); the mean can then no longer be profiled out in closed form, so a joint numerical optimisation over $(m,P)$ is required.

\begin{remark}[Computational Cost]
Each evaluation of \eqref{eq: ukf mu}--\eqref{eq: ukf Sigma} costs $O(d^3)$.
Further, re-evaluation is required at every candidate $(m,P)$, which is in contrast to \emph{\mUKFPrO{}} (\Cref{tab:complexity}; \Cref{subsec: full-update-experiment}).
However, unlike \emph{\mPrO{}} and \emph{\mUKFPrO{}}, \emph{\mUKFPrOFull{}} requires no Jacobian of $h$ to be available, which may be a practical advantage when $h$'s Jacobian is expensive or unavailable in closed form.
\end{remark}

\begin{remark}[Agreement in the Linear Case]
\label{rem:linear}
\mUKFPrO{} reduces exactly to \mPrO{} when $f$ is linear, and \mUKFPrOFull{} does so when $f$ and $h$ are both linear, by exactness of the \ac{UT} for affine maps; cf. \eqref{eq: UT moments}. Moreover, when $h$ is linear (and $f$ arbitrary), \mUKFPrOFull{} and \mUKFPrO{} minimise the same objective, so they differ only through the optimiser. The same holds for \mCKFPrO{} and \mCKFPrOFull.
\end{remark}

\section{Empirical Results}
\label{sec: experiments}

\begin{figure*}[t!]
\centering
\includegraphics[width=\textwidth]{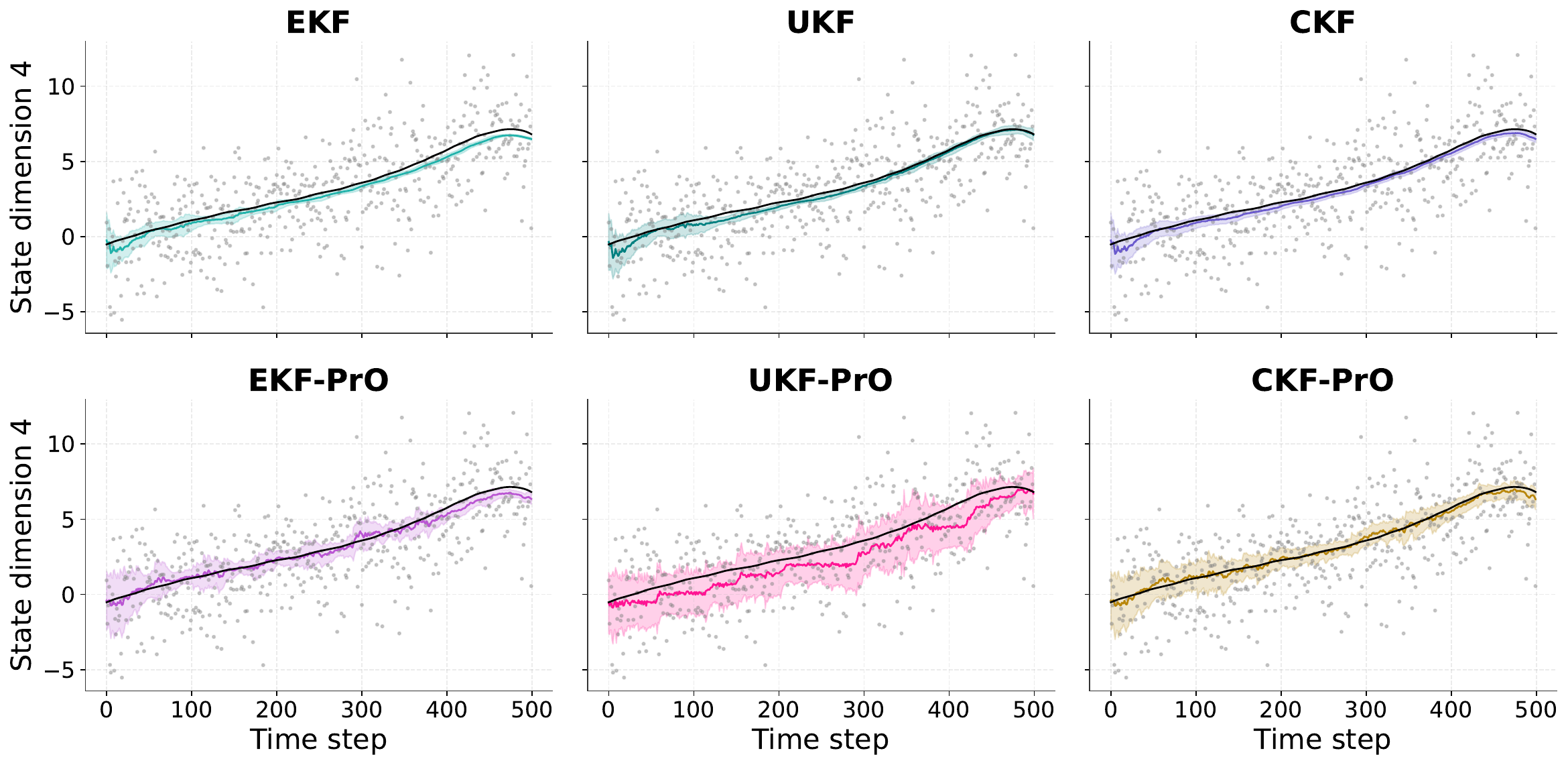}
\caption{True trajectory (black), observations (grey), filtering mean (coloured line), and $\pm 2$ standard deviation band (shaded), for one state dimension, strong-nonlinearity condition, seed 1. Panels: \mkfExtended, \mUKF, \mCKF{} (top); \mPrO, \mUKFPrO, \mCKFPrO{} (bottom).}
\label{fig:illustration}
\end{figure*}

\Cref{fig:illustration} illustrates the filtering task considered throughout this section. We evaluate \mUKFPrO{}, with both the default and the cubature parameters, on the Lorenz96 model under conditions of increasing difficulty (\Cref{subsec: Lorenz}), and the full update-step extension separately (\Cref{subsec: full-update-experiment}). Code to reproduce our experiments can be found at: \hyperlink{https://github.com/dankeann/pro-kalman-filter}{https://github.com/dankeann/pro-kalman-filter}

\subsection{Methods}
\label{sec: baselines}

We compare \mkfExtended, \mUKF, \mCKF, \mPrO, \mUKFPrO{} and \mCKFPrO. \citet{shen2026predictively} showed that \mPrO{} outperforms representative robust filters (\mWangExtended, \mAgamenoniExtended, \mWlfImq, \mWlfMd) under systematic misspecification, so we do not repeat that comparison; we also omit hierarchical and particle-based methods, which typically scale poorly to high-dimensional states. All methods are hyper-parameter free; the update step costs $O(d^3)$ for the non-\ac{PrO} filters, $O(I\,d^3)$ for the profiled-mean \ac{PrO} methods and $O(J\,d^3)$ for the full update-step extensions, where $I$ and $J$ are the optimiser iteration budgets (\Cref{tab:complexity}).

Performance is measured by two metrics. Following \citet{shen2026predictively}, we report the total squared error of the filtering mean,
\[
\mathrm{MSE} = \sum_{k=1}^{T} \|x_k - m_k\|^2,
\]
summed over all $T$ time steps and $d$ state dimensions. We also report the mean negative log-likelihood of the true state under the filtering distribution,
\[
\mathrm{NLL} = -\frac{1}{T}\sum_{k=1}^{T} \log \mathcal{N}(x_k;\, m_k, P_k),
\]
including the normalising constant. MSE measures the accuracy of the filtering mean only, whereas NLL also penalises over-confidence heavily; as the aim of \ac{PrO} filtering is calibrated inference, NLL is our principal metric.

In \Cref{subsec: Lorenz} the measurement function is the identity, since in practice $f$ is typically more strongly nonlinear than $h$; nonlinear $h$ is considered in \Cref{subsec: full-update-experiment}.

\subsection{Lorenz96 Process}
\label{subsec: Lorenz}

For our empirical evaluation, we consider the classical Lorenz96 model, commonly used to simulate the atmosphere \citep{lorenz96predictability}.
We set the latent state as $d$-dimensional, $d=60$. Condition 3 is repeated at $d=10$ and $d=80$ in \Cref{app: dimension}, with the same conclusions.
Our experimental protocol is similar to \citet{roth2017enekf}; full details, including iteration budgets and hardware/software versions, are contained in \Cref{app: Lorenz}. The state-space model is defined by a discretised ODE system for $x \in \mathbb{R}^d$: $\frac{\mathrm{d}x_i}{\mathrm{d}t} = (x_{i+1}-x_{i-2})x_{i-1} - x_i + \phi_i$, $i \in 1, 2, \hdots, d$, with $f$ given by an Euler discretisation of this system at step size $\Delta$, and the identity observation function, $h(x)=x$, observed under Gaussian noise $y_t \sim \mathcal{N}(x_t, R\cdot\mathbf{I})$. To simulate model misspecification, the \emph{data-generating} process differs from the model $f$ assumed by every filter: the true $\phi$ is not fixed, but instead depends on a latent, two-state Markov chain $m_t \in \{1,2\}$ (with persistent self-transition probability $0.995$), switching between $\phi^{(1)} = 8\cdot\mathbf{1}$ and $\phi^{(2)} \sim \mathcal{N}(8\cdot\mathbf{1}, \sigma^2\mathbf{I})$, drawn once per run, with no process noise; every filter's own model of the dynamics, by contrast, assumes the fixed, non-switching $\phi^{(1)}$ with process noise $\mathcal{N}(0,\sigma^2\Delta^2\mathbf{I})$ throughout, and is therefore unaware of the true system's occasional, hidden regime changes. All filters are initialised at $\mathcal{N}(x_0,\mathbf{I})$. We consider three conditions of increasing severity, each run across 100 seeds.

\subsubsection{Condition 1: Standard Condition}
\label{subsubsec: lorenz-standard}

First we consider the noise levels used by \citet{shen2026predictively} ($\sigma=1$, $R=5$).

\Cref{fig:lorenz-standard-mse,fig:lorenz-standard-nll} show a consistent and substantial improvement of \ac{PrO}-based methods against their classical counterparts, both in terms of \ac{MSE} and \ac{NLL}.
In this setting the performance of all \ac{PrO}-based methods is broadly similar, with a slight advantage for \mCKFPrO. 
Mean wall-clock time for \mUKFPrO{} and \mCKFPrO{} is comparable to \mPrO{} (\Cref{app: Lorenz}). 

\subsubsection{Condition 2: Increased noise}
\label{subsubsec: lorenz-wide}

As in the standard condition, but with substantially higher measurement noise and assumed process noise ($R=50$, $\sigma=3$), which also triples the size of the regime shift, stressing the predict step's covariance propagation directly.

\Cref{fig:lorenz-wide-mse,fig:lorenz-wide-nll} again show a substantial improvement resulting from \ac{PrO}-based methods compared to \mkfExtended, \mUKF, \mCKF.
Amongst the \ac{PrO}-based methods we observe a clear performance ordering
\mCKFPrO (best) $\le$ \mUKFPrO $\le$ \mPrO (worst), in both \ac{MSE} and \ac{NLL}. 
The broader predictive distribution makes accurate nonlinear moment propagation more important, which limits the local linearisation used by \mPrO{}.
 Again, mean wall-clock time for \mUKFPrO{} is comparable to \mPrO{} (\Cref{app: Lorenz}), with \mCKFPrO{} slightly slower.

\subsubsection{Condition 3: Strong nonlinearity}
\label{subsubsec: lorenz-oscillating}

As in the standard condition, but both the data-generating process and the model assumed by every filter include an additional term $2\sin(10x_i)$, chosen so that its period ($2\pi/10 \approx 0.63$) is comparable to the width of the filtering distribution, so that $f$ is genuinely curved over the region where the filter places its mass -- a regime in which a single Jacobian evaluation is a poor description of $f$.

\begin{figure}[t!]
\centering
\includegraphics[width=\columnwidth]{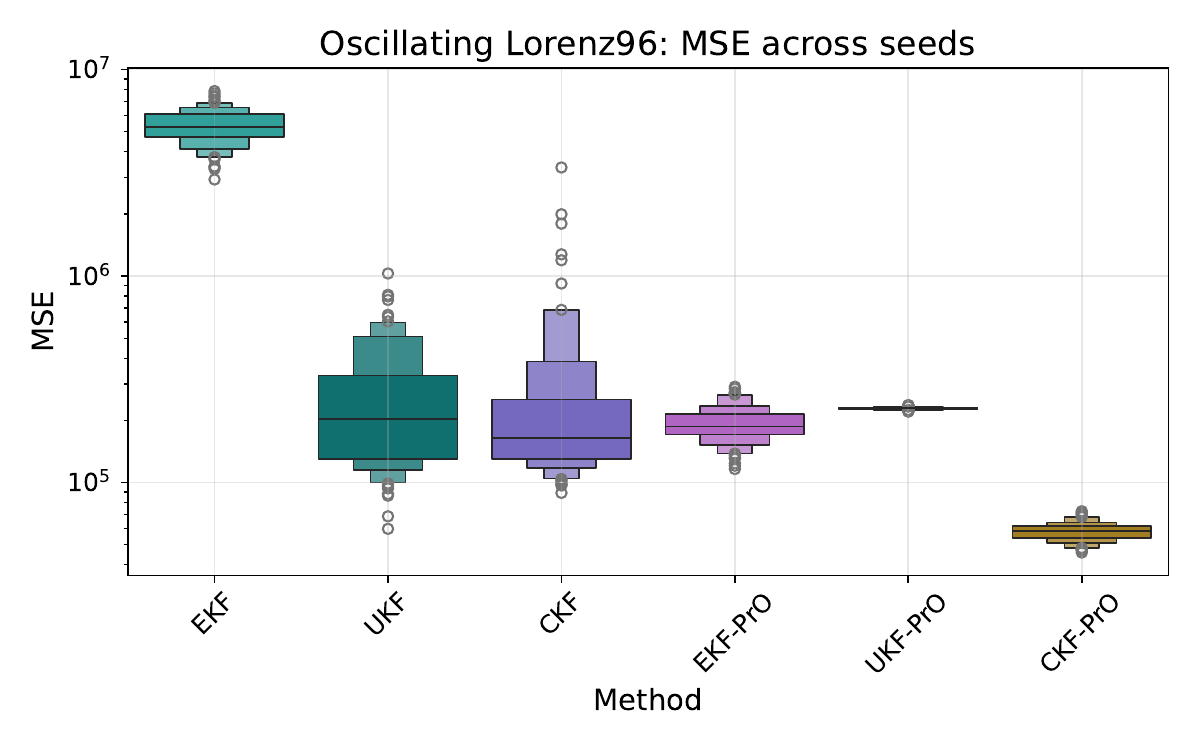}
\caption{\ac{MSE}, across 100 seeds, Condition 3 (strong nonlinearity).}
\label{fig:lorenz-oscillating-mse}
\end{figure}

\Cref{fig:lorenz-oscillating-mse,fig:lorenz-oscillating-nll} show that \mCKFPrO's \ac{MSE} and \ac{NLL} are both substantially lower than \mPrO's under this condition, in contrast to the modest differences under the conditions previously considered: \mPrO's mean \ac{MSE} is approximately $3.3\times$ and its mean \ac{NLL} approximately $27\times$ that of \mCKFPrO, and \mCKFPrO{} attains lower \ac{MSE} and \ac{NLL} than \mPrO{} on every one of the 100 seeds.
\mPrO's Jacobian is evaluated at a single point, which can be highly unrepresentative of $f$'s behaviour over the filtering distribution when $f$ is strongly curved on that scale, leading to overconfident, poorly-calibrated predictions -- reflected directly in \mPrO's substantially higher and more variable \ac{NLL} (\Cref{fig:lorenz-oscillating-nll}). 
\mCKFPrO's sigma points, spread at $\pm\sqrt{d}$ standard deviations, instead sample $f$ across the filtering distribution and capture the additional predictive uncertainty that the nonlinearity creates. \mUKFPrO{} with the default \ac{UT} settings behaves very differently: it attains the lowest \ac{NLL} of all methods but an \ac{MSE} approximately $3.9\times$ that of \mCKFPrO, because its tightly clustered sigma points ($\alpha=10^{-3}$) read the local curvature of $f$ at the mean and extrapolate it across the whole distribution, inflating the predictive covariance (\Cref{fig:lorenz-oscillating-mse}).

\mUKFPrO{} is also approximately $3$--$4\times$ slower than \mPrO{} and \mCKFPrO{} here (\Cref{fig:lorenz-oscillating-timing}); as it shares an identical update-step implementation with \mCKFPrO, we attribute this to its poorly scaled predictive covariance making the update-step optimisation harder.

\subsection{The Full Update-Step Extension}
\label{subsec: full-update-experiment}

We separately evaluate \mCKFPrOFull{} against \mCKFPrO{} on Lorenz96, on a smaller scale than \Cref{subsec: Lorenz} given the substantially higher cost of the joint $(m,P)$ optimisation underlying \mCKFPrOFull{} (\Cref{subsec: full ukf pro}). 
Since the two methods minimise the same objective when $h$ is linear, any benefit of the update-step substitution must come from nonlinearity in $h$; we therefore use the nonlinear observation function $h(x)_i = x_i + (a/\omega)\sin(\omega x_i)$ with $a=0.9$ and $\omega=10$, which is invertible for $a<1$, under the dynamics and noise levels of the standard condition (\Cref{subsubsec: lorenz-standard}). Given the cost of \mCKFPrOFull{} (\Cref{tab:timing}), we use $d=10$, $N=1{,}000$ steps and 10 seeds, with an iteration budget of $J=500$ for \mCKFPrOFull{} and $I=50$ for \mCKFPrO{}.

\mCKFPrOFull{} did not improve on \mCKFPrO{} (\Cref{tab:full-update-comparison}): its mean MSE was $1038.8$ against $825.6$, higher on 9 of the 10 seeds, while its mean \ac{NLL} was comparable ($36.1$ against $38.5$; lower on 4 of the 10 seeds), and it required approximately $7\times$ the wall-clock time (\Cref{tab:timing}), owing to the joint optimisation over $(m,P)$, in which every evaluation of the objective re-forms and propagates the sigma points. On the tests conducted here we therefore recommend \mCKFPrO{} over \mCKFPrOFull{}, although the nonlinear component of $h$ is small relative to the observation noise in this setting, and more strongly nonlinear observation models remain to be investigated.

\section{Discussion}
\label{sec: discuss}
Our central finding is that the value of replacing \mPrO's linearisation with the \ac{UT} depends critically on \emph{where} the substitution is made, and on the choice of \ac{UT} parameters. Replacing the linearisation in the predict step alone, \mUKFPrO, requires no change to \mPrO's update-step machinery, and with the cubature parameters (\mCKFPrO) yields consistent, mechanistically-understood improvements: the advantage is small in the standard condition (\Cref{subsubsec: lorenz-standard}), moderate in the wide-covariance condition (\Cref{subsubsec: lorenz-wide}) and very substantial under strong nonlinearity (\Cref{subsubsec: lorenz-oscillating}), where $f$ is strongly curved over the region where the filter places its mass and a single Jacobian evaluation is a poor description of it. 
With the default \ac{UT} parameters, by contrast, \mUKFPrO{} is unreliable: its tightly clustered sigma points read the local curvature of $f$ at the mean and extrapolate it across the whole distribution, which can accidentally improve calibration or severely degrade accuracy (\Cref{subsubsec: lorenz-oscillating}). In the update step, extending the same substitution to $h$ (\mUKFPrOFull, \Cref{subsec: full ukf pro}) requires abandoning the closed-form profiled-mean update that makes \mPrO efficient, and -- on the tests conducted here, with the cubature parameters (\Cref{subsec: full-update-experiment}) -- does not appear to earn its substantially higher cost.

Our work contributes to an emerging literature on \ac{PrO} posteriors \citep{mclatchie2025predictively,lai2024predictive,shenprediction,shen2026predictively}, extending their applicability to strongly nonlinear state-space models.
On the other hand, the theoretical properties of \mUKFPrO{} (e.g. statistical efficiency, dimension dependence, stability) are not yet understood, and further work will be needed to extend existing theoretical analysis of \ac{PrO} posteriors to this setting. All our experiments use the Lorenz96 model and, following \citet{shen2026predictively}, a small assumed process noise, so further benchmarks will be needed to establish how widely our conclusions hold.
The \ac{PrO} methods are roughly $I$ times slower than their non-\ac{PrO} counterparts (around $50\times$ at $I=50$), with \mCKFPrO{} costing about the same as \mPrO.

Several further extensions could be considered: higher-order sigma-point/cubature rules \citep{ito2002gaussian} could be used in place of the third-degree cubature rule if its accuracy proves insufficient; the update step's Gaussian variational family could eventually be relaxed in favour of particle-based methods analogous to the ensemble Kalman filter \citep{evensen2003ensemble}, at increased computational cost; and it remains open whether \mUKFPrOFull{} offers a benefit for observation models that are more strongly nonlinear relative to the observation noise than those considered in \Cref{subsec: full-update-experiment}.

\paragraph{Acknowledgments}
CJO was supported by a Philip Leverhulme Prize (PLP-2023-004).

\bibliographystyle{abbrvnat}
\bibliography{bibliography}

\appendix
\section*{Supplementary Material}

This document contains supplementary material for the paper \emph{An Unscented PrO Kalman Filter}.
\Cref{app: EKF PrO detail} completes our discussion of earlier work by presenting the precise local linearisation used in the \mPrO of \citet{shen2026predictively}.
\Cref{app: protocol} contains full details required to reproduce the empirical results in the main text.
\Cref{app: additional results} collects additional figures and tables referred to in the main text.

\section{Implementational Details}
\label{app: EKF PrO detail}

\citet{shen2026predictively} extend \mPrOKF to nonlinear $f, h$ by linearising both, exactly as the \mkfExtended does: the prediction step linearises $f$ around $m_{k-1}$ to obtain $\tilde m_k, \tilde P_k$ via the usual Jacobian sandwich formula, and the update step linearises $h$ around the current estimate of the filtering mean $m_k^\star(P)$ to obtain
\begin{align}
    S_k(P) &= \mathbf{H}_{x}(m_k^\star, 0)\: P\:\mathbf{H}_{x}(m_k^\star, 0)^\top + \mathbf{H}_\xi(m_k^\star, 0)\: R_k\: \mathbf{H}_\xi(m_k^\star, 0)^\top, \label{eq: pro linear1}\\
    v_k(m) &= y_k - h(m_k^\star, 0) - \mathbf{H}_{x}(m_k^\star, 0)(m-m_k^\star) \label{eq: pro linear2},
\end{align}
which is then plugged directly into \eqref{eq: pro_gaussian}; this is \mPrO, analogous to the \emph{iterative} \mkfExtended (\citealp{gelb1974applied}; Algorithm~7.9 in \citealp{Sarkka23}).
All profiled-mean methods (\mPrO, \mUKFPrO, \mCKFPrO) share an identical update-step implementation, which minimises the profiled objective over a Cholesky parametrisation of $P$, initialised at the prior, using L-BFGS.

\section{Experimental Protocol}
\label{app: protocol}

\paragraph{Hardware Requirement}
Experiments were run locally on a MacBook Pro (2.3 GHz 8-Core Intel Core i9, 32 GB 2667 MHz DDR4 memory) using JAX (CPU backend), macOS 15.7.9, JAX version 0.4.38, NumPy version 2.3.5, Python 3.11.5.

\paragraph{JAX Implementation.}
Our experimental code is implemented in JAX. The code framework builds on the
third-party package \texttt{rebayes\_mini} (\url{https://github.com/gerdm/rebayes-mini/tree/main}), which provides a
lightweight implementation framework for recursive Bayesian filtering methods.
We use this package for shared filtering abstractions and baseline
implementations, and implement the method-specific updates, experimental
configurations, and evaluation scripts used in this paper on top of this
framework. All wall-clock times include JIT compilation, which is repeated for every seed, and are measured after all computation has completed.

\subsection{Lorenz96 Process}
\label{app: Lorenz}
The dynamics, data-generating process, filter model and initialisation are as described in \Cref{subsec: Lorenz}, with $\Delta = 0.002$, $d=60$ unless stated otherwise, and $N = 10{,}000$ steps per seed over seeds $1,\dots,100$. Condition~1 uses $(\sigma, R) = (1, 5)$; Condition~2 uses $(3, 50)$; Condition~3 uses $(1, 5)$ with the additional term $2\sin(10x_i)$ added to the right-hand side of the dynamics in both the data-generating process and the filter model. MSE and NLL are computed as described in \Cref{sec: baselines}.

\begin{figure}[htbp]
\centering
\begin{subfigure}{0.62\textwidth}
    \centering
    \includegraphics[width=\linewidth]{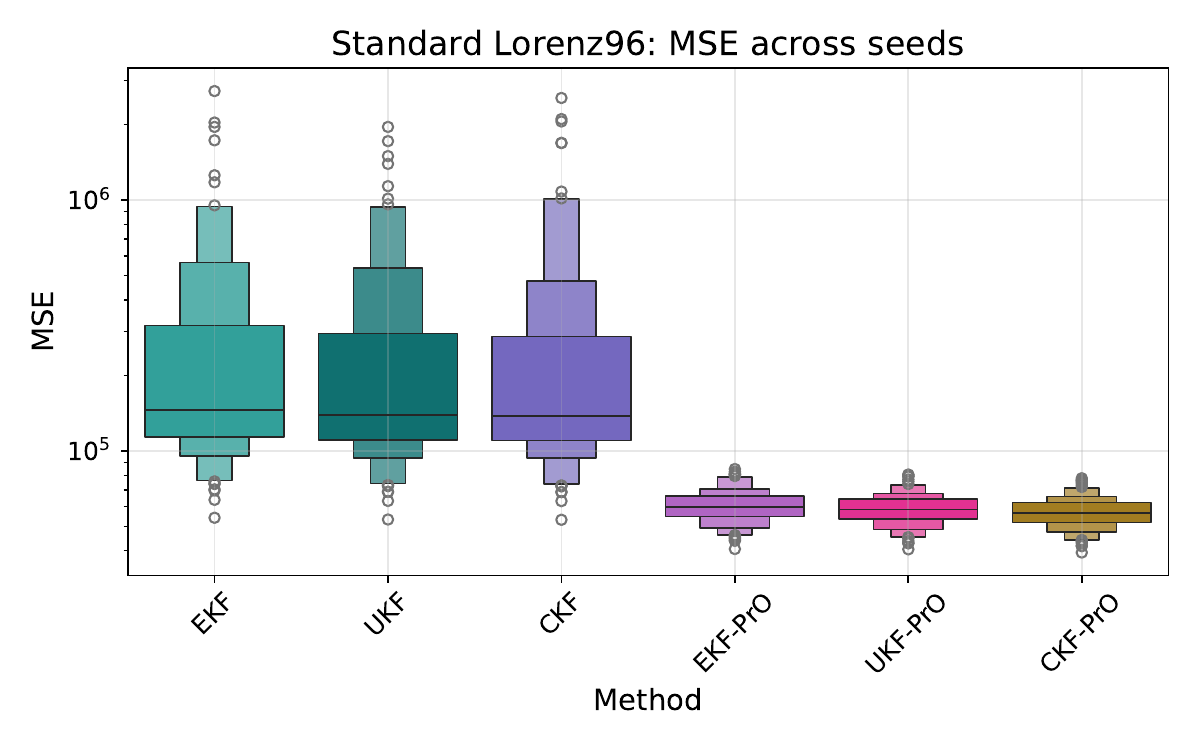}
    \caption{\ac{MSE}.}
    \label{fig:lorenz-standard-mse}
\end{subfigure}

\vspace{0.6em}
\begin{subfigure}{0.95\textwidth}
    \centering
    \includegraphics[width=\linewidth]{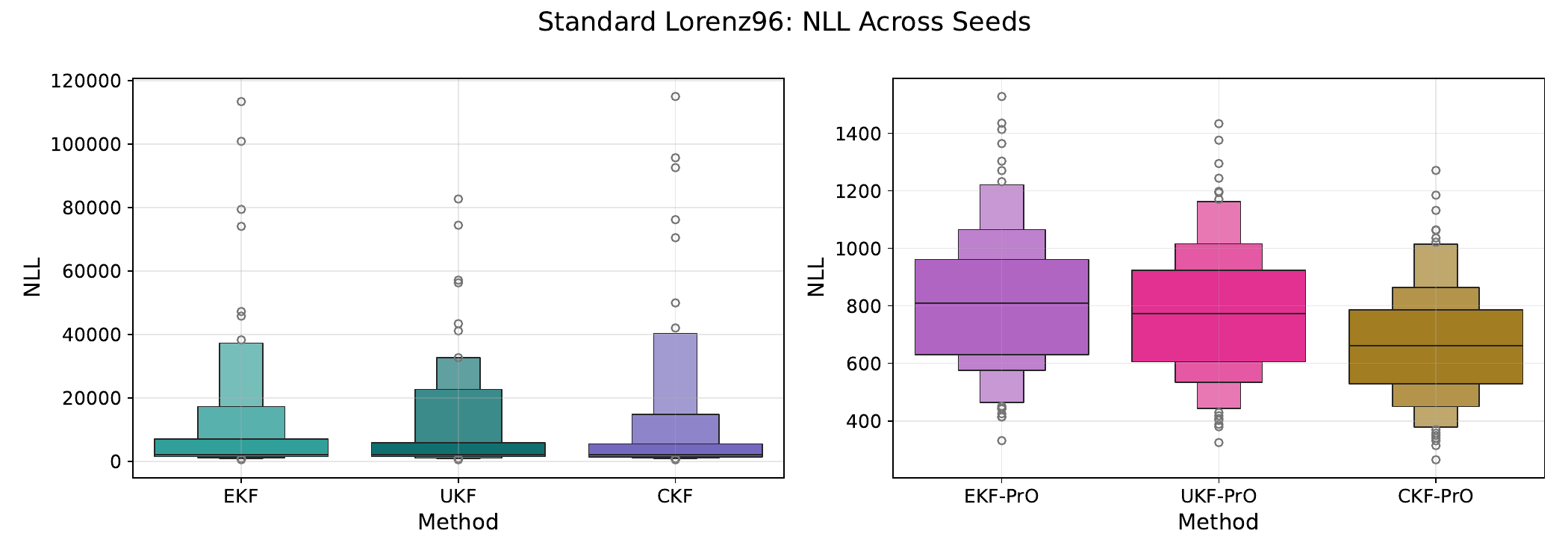}
    \caption{Mean \ac{NLL}, shown separately for the plain filters (left) and \ac{PrO} variants (right) given the difference in scale.}
    \label{fig:lorenz-standard-nll}
\end{subfigure}
\caption{Results across 100 seeds, Condition 1 (standard).}
\label{fig:lorenz-standard}
\end{figure}

\begin{figure}[htbp]
\centering
\begin{subfigure}{0.62\textwidth}
    \centering
    \includegraphics[width=\linewidth]{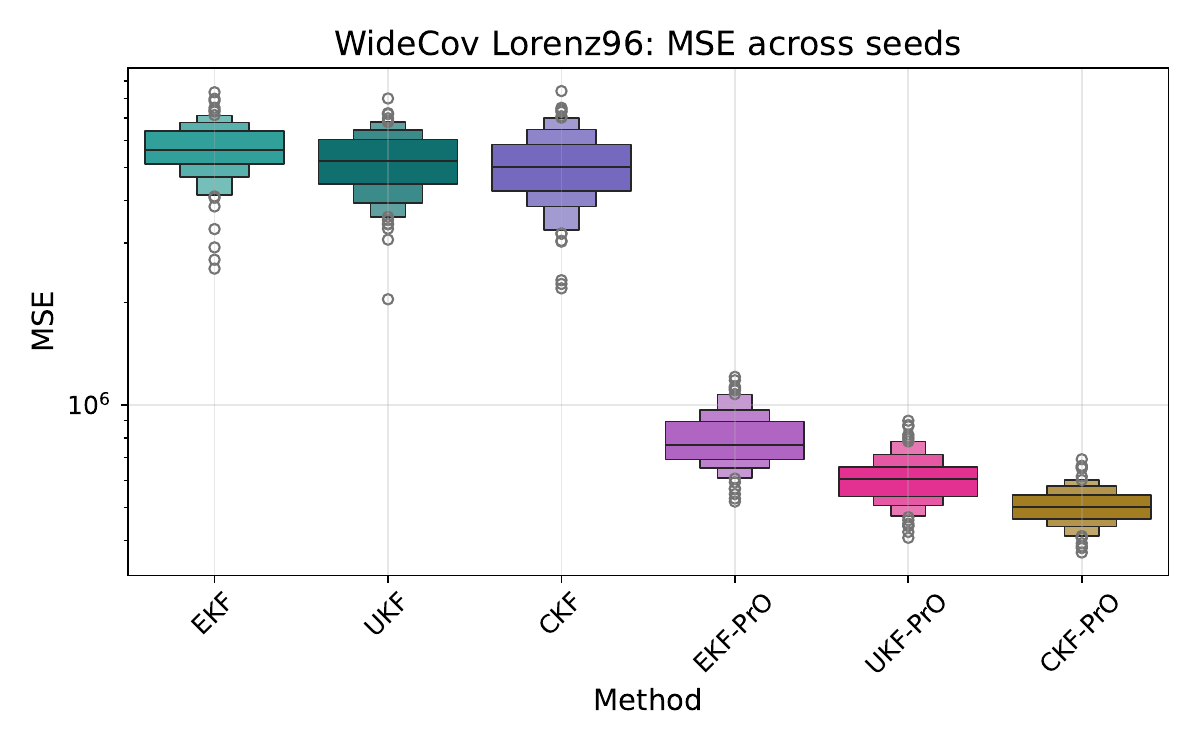}
    \caption{\ac{MSE}.}
    \label{fig:lorenz-wide-mse}
\end{subfigure}

\vspace{0.6em}
\begin{subfigure}{0.95\textwidth}
    \centering
    \includegraphics[width=\linewidth]{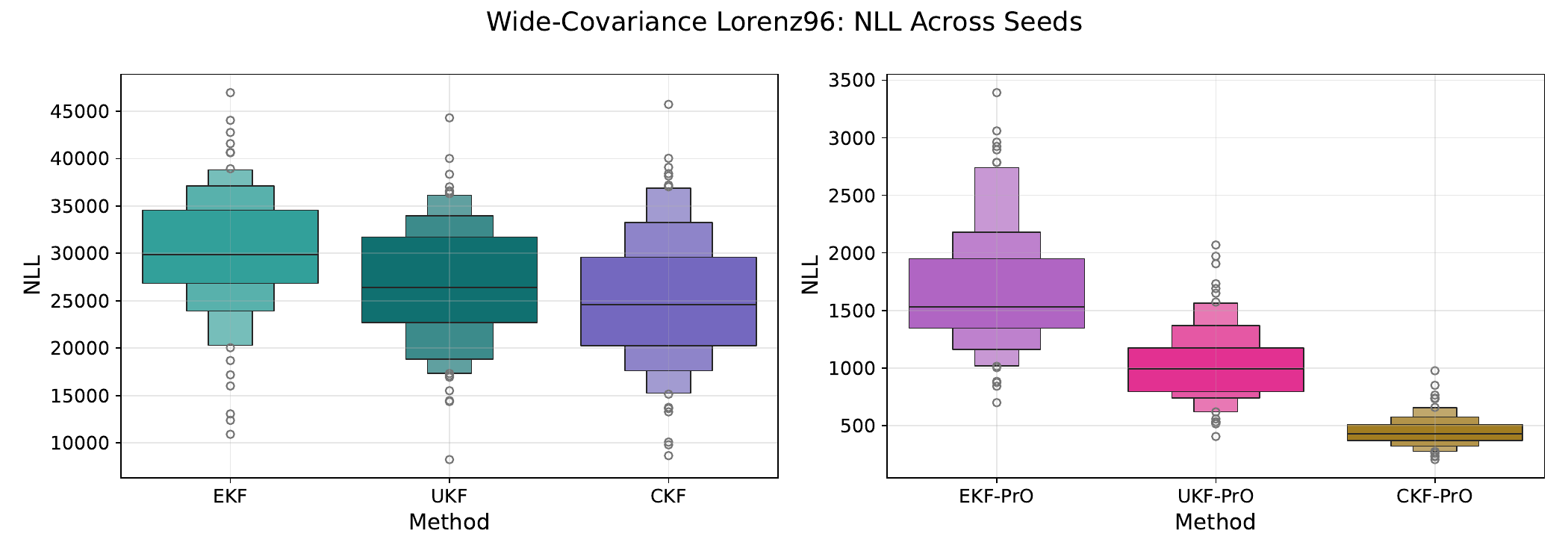}
    \caption{Mean \ac{NLL}, shown separately for the plain filters (left) and \ac{PrO} variants (right) given the difference in scale.}
    \label{fig:lorenz-wide-nll}
\end{subfigure}
\caption{Results across 100 seeds, Condition 2 (wide covariance).}
\label{fig:lorenz-wide}
\end{figure}

\begin{figure}[htbp]
\centering
\begin{subfigure}{0.95\textwidth}
    \centering
    \includegraphics[width=\linewidth]{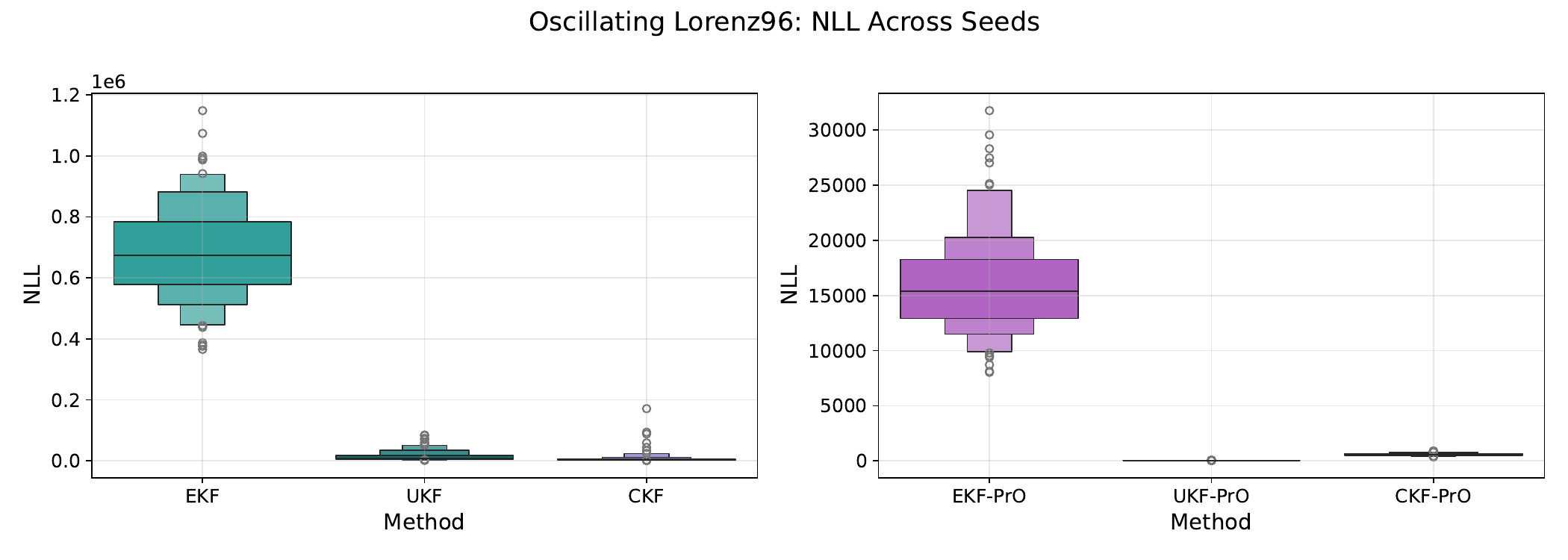}
    \caption{Mean \ac{NLL}, shown separately for the plain filters (left) and \ac{PrO} variants (right) given the difference in scale.}
    \label{fig:lorenz-oscillating-nll}
\end{subfigure}

\vspace{0.6em}
\begin{subfigure}{0.55\textwidth}
    \centering
    \includegraphics[width=\linewidth]{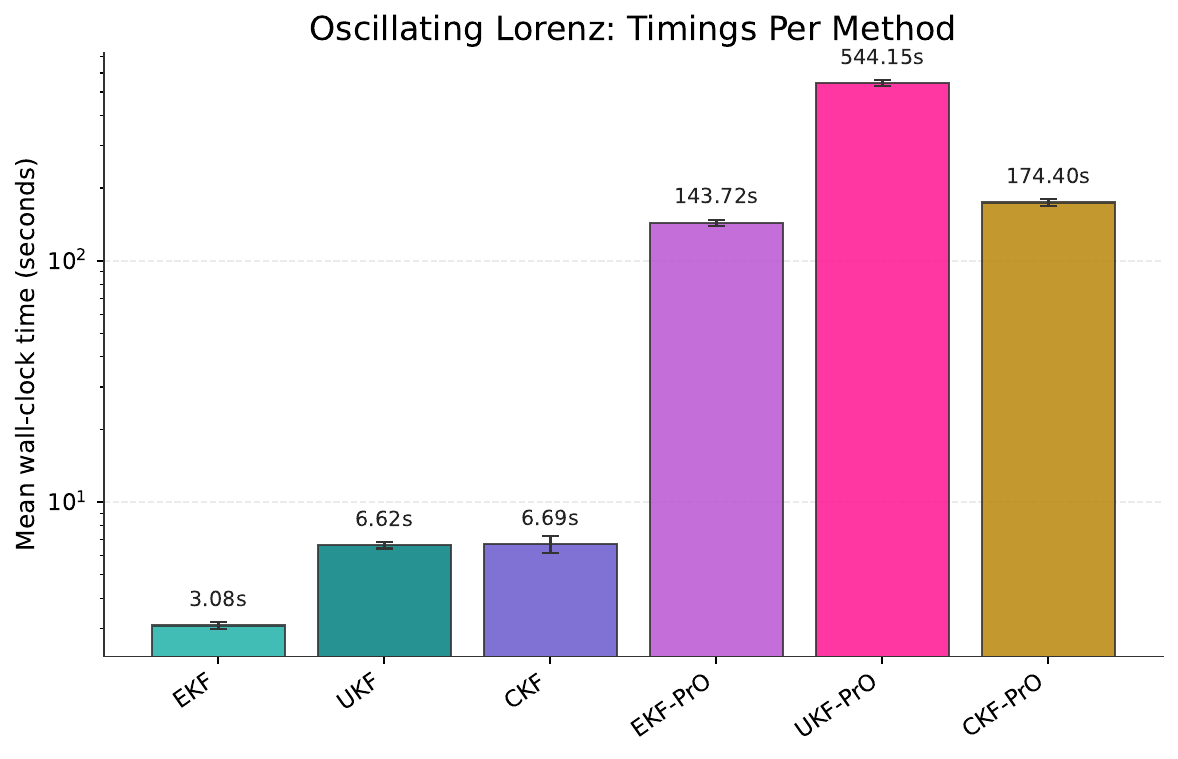}
    \caption{Mean wall-clock time per seed, including JIT compilation. \mUKFPrO{} is approximately $3$--$4\times$ slower than \mPrO{} and \mCKFPrO{} under this condition specifically.}
    \label{fig:lorenz-oscillating-timing}
\end{subfigure}
\caption{Results across 100 seeds, Condition 3 (strong nonlinearity). The corresponding \ac{MSE} is shown in \Cref{fig:lorenz-oscillating-mse}.}
\label{fig:lorenz-oscillating-extra}
\end{figure}

For \mPrO, \mUKFPrO{} and \mCKFPrO{} across all three conditions, we use $n_{\mathrm{inner}}=5$ and $n_{\mathrm{outer}}=10$, i.e.\ at most $I=50$ inner iterations per step.
\Cref{fig:lorenz-standard-mse,fig:lorenz-standard-nll,fig:lorenz-wide-mse,fig:lorenz-wide-nll,fig:lorenz-oscillating-mse,fig:lorenz-oscillating-nll,fig:lorenz-oscillating-timing} show the results for all three conditions, and \Cref{tab:timings} reports wall-clock times.

For the comparison of \mCKFPrO{} and \mCKFPrOFull{} (\Cref{subsec: full-update-experiment}), we use the dynamics and noise levels of Condition~1 with the observation function $h(x)_i = x_i + (0.9/10)\sin(10 x_i)$, $d=10$, $N=1{,}000$ steps and seeds $1,\dots,10$. \mCKFPrOFull's joint optimisation over $(m,P)$ has no separate outer/inner structure (\Cref{subsec: full ukf pro}); we use a total budget of $J=500$ iterations ($n_{\mathrm{inner}}=50$, $n_{\mathrm{outer}}=10$, retained purely as a naming convention), against $I=50$ for \mCKFPrO.

\begin{table}[h]
    \small
    \centering
    \begin{tabular}{lccc}
    \toprule
    Method & Condition 1 & Condition 2 & Condition 3 \\
    \midrule
    \mkfExtended & $2.79 \pm 0.14$   & $2.80 \pm 0.07$   & $3.08 \pm 0.11$ \\
    \mUKF        & $5.21 \pm 0.18$   & $5.36 \pm 0.32$   & $6.62 \pm 0.19$ \\
    \mCKF        & $5.33 \pm 0.43$   & $5.34 \pm 0.28$   & $6.69 \pm 0.53$ \\
    \mPrO        & $157.85 \pm 4.73$ & $157.82 \pm 2.18$ & $143.72 \pm 4.04$ \\
    \mUKFPrO     & $159.63 \pm 5.47$ & $158.94 \pm 3.39$ & $544.15 \pm 15.28$ \\
    \mCKFPrO     & $159.78 \pm 4.25$ & $168.73 \pm 3.18$ & $174.40 \pm 5.15$ \\
    \bottomrule
    \end{tabular}
    \caption{Mean wall-clock time per seed in seconds (mean $\pm$ std over 100 seeds), $d=60$, $N=10{,}000$, including JIT compilation.}
    \label{tab:timings}
\end{table}

\subsection{Robustness to State Dimension}
\label{app: dimension}
To check that our findings are not specific to $d=60$, we repeat the strong-nonlinearity condition (\Cref{subsubsec: lorenz-oscillating}) at $d=10$ and $d=80$, with all other settings unchanged except that, given the computational cost, we use $N=1{,}000$ steps and $10$ seeds at both dimensions, rather than the $N=10{,}000$ steps and $100$ seeds used at $d=60$. Absolute values are therefore not directly comparable with those at $d=60$, and we compare methods within each dimension only.

\begin{table}[h]
    \small
    \centering
    \begin{tabular}{lcccc}
    \toprule
    & \multicolumn{2}{c}{$d=10$ ($N=1{,}000$, 10 seeds)} & \multicolumn{2}{c}{$d=80$ ($N=1{,}000$, 10 seeds)} \\
    Method & MSE & Mean NLL & MSE & Mean NLL \\
    \midrule
    \mkfExtended & $1647.28 \pm 936.82$ & $810.95 \pm 576.41$ & $17895.38 \pm 10527.79$ & $10656.16 \pm 6182.70$ \\
    \mUKF        & $795.95 \pm 529.61$  & $74.45 \pm 103.87$  & $6676.80 \pm 1565.81$   & $1260.61 \pm 668.60$ \\
    \mCKF        & $964.05 \pm 527.52$  & $171.56 \pm 204.55$ & $6728.96 \pm 3109.71$   & $885.42 \pm 567.86$ \\
    \mPrO        & $1332.23 \pm 736.31$ & $657.45 \pm 558.11$ & $10398.27 \pm 2927.71$  & $6328.41 \pm 2408.60$ \\
    \mUKFPrO     & $3027.13 \pm 106.79$ & $8.88 \pm 0.19$     & $52260.97 \pm 1225.01$  & $60.72 \pm 1.65$ \\
    \mCKFPrO     & $943.10 \pm 143.82$  & $44.78 \pm 50.59$   & $5654.33 \pm 978.53$    & $306.62 \pm 189.76$ \\
    \bottomrule
    \end{tabular}
    \caption{Mean $\pm$ std MSE and mean NLL, strong-nonlinearity condition, at $d=10$ and $d=80$.}
    \label{tab:dimension-robustness}
\end{table}

The same pattern holds at both dimensions: \mCKFPrO{} attains lower MSE and substantially lower \ac{NLL} than \mPrO{} (MSE ratios of approximately $1.4\times$ at $d=10$ and $1.8\times$ at $d=80$; \ac{NLL} ratios of approximately $15\times$ and $21\times$), and \mUKFPrO{} again attains the lowest \ac{NLL} but the highest MSE of the \ac{PrO} methods. At $d=10$ the plain \mUKF{} attains a slightly lower mean MSE than \mCKFPrO, but with substantially higher variability across seeds and a higher \ac{NLL}.

The relative costs are also consistent with $d=60$: \mCKFPrO's wall-clock time is $\sim 1.1\times$ \mPrO's at $d=10$ ($3.14$s vs.\ $2.79$s) and $\sim 1.4\times$ at $d=80$ ($73.61$s vs.\ $54.16$s), while \mUKFPrO's is $\sim 1.4\times$ and $\sim 2.7\times$ respectively, all including JIT compilation. We do not test beyond $d=80$ given this cost; further methodological development will be needed to realise high-dimensional \ac{PrO} filtering.

\clearpage
\section{Additional Results}
\label{app: additional results}

This section collects the figures and tables that are referred to, but not shown, in the main text.

\begin{figure}[htbp]
\centering
\begin{subfigure}{0.48\textwidth}
    \centering
    \includegraphics[width=\linewidth]{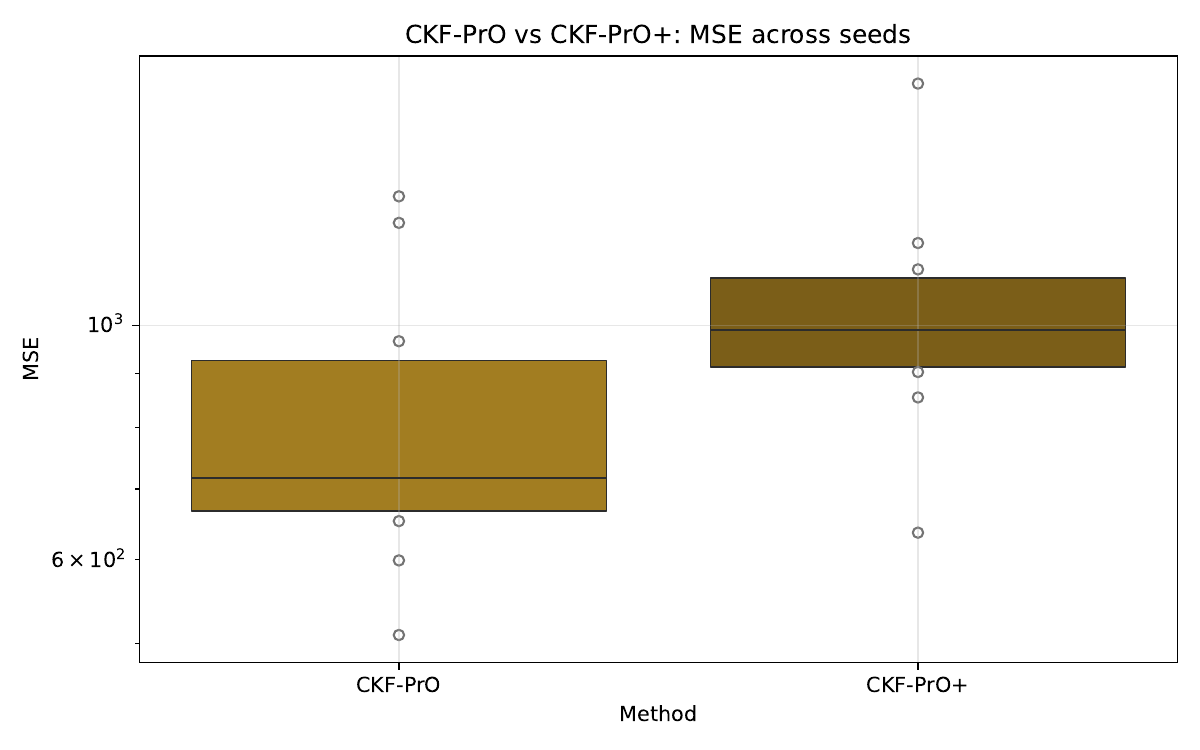}
    \caption{\ac{MSE}.}
    \label{fig:full-update-mse}
\end{subfigure}
\hfill
\begin{subfigure}{0.48\textwidth}
    \centering
    \includegraphics[width=\linewidth]{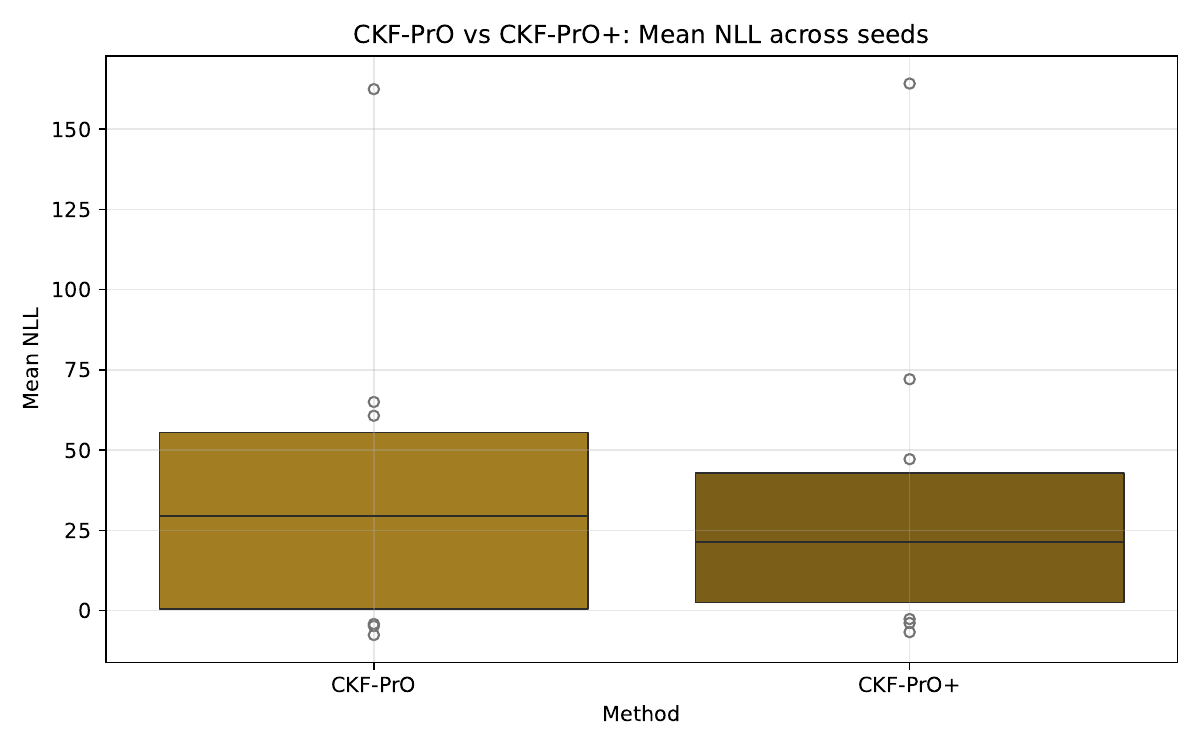}
    \caption{Mean \ac{NLL}.}
    \label{fig:full-update-nll}
\end{subfigure}
\caption{\mCKFPrO{} and \mCKFPrOFull{}, nonlinear $h$, standard condition, $d=10$, $N=1{,}000$, 10 seeds.}
\label{fig:full-update}
\end{figure}

\begin{table}[htbp]
\centering
\small
\begin{minipage}[t]{0.55\textwidth}
    \centering
    \begin{tabular}{lcc}
    \toprule
    Method & MSE & \ac{NLL} \\
    \midrule
    \mCKFPrO     & $825.63 \pm 258.63$  & $38.47 \pm 48.19$ \\
    \mCKFPrOFull & $1038.76 \pm 263.52$ & $36.13 \pm 48.53$ \\
    \bottomrule
    \end{tabular}
    \caption{\mCKFPrO{} and \mCKFPrOFull{}, nonlinear $h$, $d=10$, $N=1{,}000$, mean $\pm$ std over 10 seeds.}
    \label{tab:full-update-comparison}
\end{minipage}
\hfill
\begin{minipage}[t]{0.40\textwidth}
    \centering
    \begin{tabular}{lc}
    \toprule
    Method & Wall-clock time \\
    \midrule
    \mCKFPrO     & $8.02 \pm 0.84$s\\
    \mCKFPrOFull & $57.34 \pm 3.75$s\\
    \bottomrule
    \end{tabular}
    \caption{Mean wall-clock time per seed, same setting, including JIT compilation.}
    \label{tab:timing}
\end{minipage}
\end{table}

\begin{table}[htbp]
    \small
    \centering
    \begin{tabular}{llll}
    \toprule
               Method & Cost & \#HP & Ref \\
    \midrule
         \mkfExtended &  $O(d^3)$ & 0 & \citet{kalman1960}\\
         \mUKF &  $O(d^3)$ & 0 & \citet{julier1996general}\\
         \mCKF &  $O(d^3)$ & 0 & \citet{arasaratnam2009}\\
         \mPrO & $O(I\,d^3)$ & 0 & \citet{shen2026predictively} \\
         \mUKFPrO & $O(I\,d^3)$ & 0 & (Ours) \\
         \mCKFPrO & $O(I\,d^3)$ & 0 & (Ours) \\
         \mUKFPrOFull & $O(J\,d^3)$ & 0 & (Ours) \\
         \mCKFPrOFull & $O(J\,d^3)$ & 0 & (Ours) \\
    \bottomrule
    \end{tabular}
    \caption{
        Computational complexity of the update step for each method used in our experiments.
        Here $d$ is the dimension of the latent state, $I$ is the number of optimiser iterations for the profiled-mean methods, and $J$ is the number of iterations for the jointly-optimised full-update methods ($I=50$ and $J=500$ in our experiments).
        \#HP is the number of hyper-parameters to be specified (in this paper only hyper-parameter free methods are considered).
        }
    \label{tab:complexity}
\end{table}

\end{document}